# Calculation of the Local Standard of Rest from 20 574 Local Stars in the New Hipparcos Reduction with Known Radial Velocities

## Charles Francis[1], Erik Anderson[2]

[1] 25 Elphinstone Rd., Hastings, TN34 2EG, UK.
[2] 800 Morton St., Ashland, OR 97520, USA.

## ARTICLE INFO

## ABSTRACT

*Context.* An accurate estimate of the local standard of rest (LSR) is required to determine key parameters used in approximate galactic mass models and to understand Galactic structure and evolution. However, authors are often forced to base dynamical analyses on potentially unreliable figures because recent determinations of the LSR have failed to reach agreement, especially with regard to the direction, $V$, of Galactic rotation.
*Aims.* To explain why the traditional method for calculating the LSR fails, and to find alternative means of calculating the LSR with realistic error margins.
*Methods.* We assemble and investigate the kinematic properties of 20 574 stars within 300 pc, with complete and accurate kinematic data. The traditional method of calculating the LSR assumes a well-mixed distribution. In fact, the velocity distribution is highly structured, invalidating calculations based on mean motions and asymmetric drift. We find other indicators in the distribution which we believe give a better estimate of circular motion.
*Results.* We find good agreement between results and give as our best estimate of the LSR $(U_0, V_0, W_0)$ = $(7.5 \pm 1.0, 13.5 \pm 0.3, 6.8 \pm 0.1)$ km s$^{-1}$. We calculate the slope of the circular speed curve at the solar radius, finding -9.3 $\pm$0.9 km s$^{-1}$kpc$^{-1}$.

## 1  Introduction

The local standard of rest (LSR) is defined to mean the velocity of a circular orbit at the Solar radius from the Galactic centre. The definition idealizes an axisymmetric galaxy in equilibrium, ignoring features like the bar, spiral arms, and perturbations due to satellites. An accurate estimate of the LSR is required to determine parameters like the enclosed mass at the solar radius for use in approximate mass models (e.g., Klypin, Zhao & Somerville, 2002) and the eccentricity distribution which is of importance in understanding Galactic structure and evolution. In the absence of a rigorous determination of the LSR, authors are often forced to base dynamical analyses on a potentially unreliable figure. Recent determinations of the LSR, using different stellar populations and minor differences in methodology, (table 1) have failed to reach agreement, especially with regard to the direction, $V$, of Galactic rotation. Quoted error margins are much less than the variation in the figures found for the LSR. A number of factors may contribute to this, including lack of complete kinematic information, selection criteria, and the irregular velocity distribution of the population due to bulk motions – either moving groups formed from particular gas clouds or streams arising from large scale dynamics (Dehnen, 1998; Fux, 2001; Famaey et al., 2005; Chakrabarty, 2007; Klement et al, 2008).

The usual way to calculate the LSR is to calculate the mean velocity of a stellar population, and to correct for asymmetric drift (e.g., Binney & Tremaine, 1987, pp. 198-9). This method requires a well-mixed distribution and is vulnerable to kinematic bias due to bulk motions within star populations. Usually it is hoped that in an average, over many streams each representing only a minor fraction of the whole sample, the effect of bulk motions will largely average out. This assumption is not borne out in the data. There have been a number of recent studies challenging the assumption of a well mixed distribution (Skuljan, Hearnshaw & Cottrell, 1999; Fux 2001; Dehnen, 1998; Famaey et al., 2005, Chakrabarty, 2004 & 2007, Quillen 2003, Quillen & Minchev 2006, de Simone et al, 2004, Chakrabarty & Sideris, 2008).

It is therefore appropriate to review the calculation of the LSR and to consider how the structure of the distribution affects the result. After removal of fast moving stars, a few streams contribute over one third of the entire population (Famaey et al., 2005), and inevitably bias a traditional analysis. Moreover, the composition of the streams is strongly dependent on both age and colour (Dehnen, 1998). We will show that this seriously affects calculations of the asymmetric drift from the velocity and dispersion of different populations. We will conclude that the principal reason for disagreement between the analyses tabulated in table 1 is that the influence of streams has not been understood. It is well known in statistics that an inhomogeneous population can lead to errors in analysis (e.g. Bissantz & Munk, 2001, 2002). We will consider other indicators based on a large population of single stars and solved spectroscopic binaries with complete kinematic data. We find good agreement between indicators.

In section 2 we describe our stellar population, culled from published data bases, and discuss effects of possible kinematic bias. In section 3 we determine a velocity ellipsoid using Gaussian fitting. In section 4 we consider Parenago's discontinuity and show that it is eliminated for stars inside the velocity ellipsoid. In section 5 we describe traditional calculations of the LSR and find that results are not consistent and in some cases not reasonable, since they indicate a disproportionately large number of stars on the outer part of their orbit. In sections 6 and 7 we describe the structure of the velocity distribution, showing how dependencies on colour and age invalidate traditional calculations of the LSR and cast light on the nature of Parenago's discontinuity. In section 8 we describe the eccentricity vector, and in section 9 we replace the velocity ellipsoid with a cut on eccentricity, and use Gaussian fitting to find a value for the LSR which needs no further correction for asymmetric drift. In section 10 we argue that an observed minimum in the velocity distributions is also an indicator of the LSR, and calculate a value in agreement with the previous estimate. In section 11 we show that this minimum also gives a method of calculating the circular speed curve avoiding bias due to particular streaming motions. Our conclusions are summarized in section 12.

| Source | Notes | Data | $U_0$ | $V_0$ | $W_0$ |
|--------|-------|------|-------|-------|-------|
| Bobylev & Bajkova (2007) | F & G dwarfs | 3D motions | 8.7 ± 0.5 | 6.2 ± 2.2 | 7.2 ± 0.8 |
| Bobylev, et al. (2006) | from the PCRV | 3D motions | 10.2 ± 0.4 | 10.9 ± 0.4 | 6.6 ± 0.4 |
| Hogg, et al. (2005) | | proper motions | 10.1 ± 0.5 | 4.0 ± 0.8 | 6.7 ± 0.2 |
| Fehrenbach, et al. (2001) | Avg. Dist. = 46 pc | radial velocities | 9.79 ± 0.5 | 13.20 ± 0.5 | 3.25 ± 0.9 |
| Fehrenbach, et al. (2001) | Avg. Dist. = 195 pc | radial velocities | 8.24 ± 0.6 | 11.58 ± 0.6 | 5.97 ± 1.1 |
| Fehrenbach, et al. (2001) | Avg. Dist. = 378 pc | radial velocities | 2.93 ± 0.6 | 10.36 ± 0.6 | 4.79 ± 1.2 |
| Mignard (1999) | A0-F5, 100pc - 2kpc | proper motions | 11.0 | 10.87 | 7.23 |
| Mignard (1999) | K0-K5 | proper motions | 9.88 | 14.19 | 7.76 |
| Miyamoto and Zhu (1998) | 159 Hipparcos Cepheids | proper motions | 10.62 ± 0.49 | 16.06 ± 1.14 | 8.60 ± 1.02 |
| Dehnen & Binney (1998) | Max Dist. ~ 100pc, Hipparcos | proper motions | 10.00 ± 0.36 | 5.23 ± 0.62 | 7.17 ± 0.38 |
| Binney et al. (1997) | Stars near south celestial pole | proper motions | 11 ±0.6 | 5.3±1.7 | 7.0±0.6 |
| Jaschek, et al. (1991a) | Mean, Bright Star Catalogue | radial velocities | 11.4 | 14.7 | 7.6 |
| Jaschek, et al. (1991b) | Median | radial velocities | 9.8 | 11.6 | 5.9 |
| Jaschek, et al. (1991b) | Mode | radial velocities | 8.6 | 7.2 | 3.8 |
| Mihalas & Binney (1981) | Galactic astronomy: 2nd ed. | | 9.2 ± 0.3 | 12.0 | 6.9 ±0.2 |
| Mayor (1974) | A & F stars | 3D motions | 10.3 ± 1 | 6.3 ± 0.9 | 5.9 ± 0.4 |

**Table 1:** Recent measurements of the LSR fail to converge, particularly in the *V*-direction.

## 2 Our Sample

### 2.1 Stellar Databases

To minimize the influence of random errors on results, it is important to use stars for which accurate measurement is available. Hipparcos provided parallax measurements of unsurpassed accuracy. Systematic parallax errors are stated at less than 0.1 mas (ESA, 1997), or less than 3% for a star at 300pc. We derived a stellar population with kinematically complete data by combining astrometric parameters from the recently released catalogue, *Hipparcos, the New Reduction of the Raw Data* (van Leeuwen, 2007a; hereafter "HNR") plus the Tycho-2 catalogue (ESA, 1997) with radial velocity measurements contained in the *Second Catalogue of Radial Velocities with Astrometric Data* (Kharchenko, et al., 2007; hereafter "CRVAD-2").

HNR claims improved accuracy by a factor of up to 4 over the original Hipparcos catalogue (ESA, 1997) for nearly all stars brighter than magnitude 8. The improvement is due to the increase of available computer power since the original calculations from the raw data, to an improved understanding of the Hipparcos methodology, which compared positions of individual stars to the global distribution and incorrectly weighted stars in high-density star fields leading to the well-known 10% error in distance to the Pleiades, and to better understanding of noise, such as dust hits and scanphase jumps. *Validation of the New Hipparcos Reduction* (van Leeuwen, 2007b) "confirms an improvement by a factor 2.2 in the total weight compared to the catalogue published in 1997, and provides much improved data for a wide range of studies on stellar luminosities and local galactic kinematics." Our analysis showed evidence of the improvement in the data by comparison with a preliminary analysis based on the previous data set, both by substantially increasing the number of stars with parallax errors less than 20%, and by showing moving groups as sharper spikes in the velocity distribution – errors will tend to smear out such spikes.

CRVAD-2 contains most of the stars in two important radial velocity surveys: *The Geneva-Copenhagen survey of the Solar neighbourhood* (Nordström, et al., 2004; hereafter "G-CS"), which surveyed nearby F and G dwarfs, and *Local Kinematics of K and M Giants from CORAVEL* (Famaey, et al., 2005; hereafter "Famaey"). We included about 300 stars in G-CS and Famaey not given in CRVAD-2 and incorporated the revised ages for G-CS II (Holmberg, Nordström and Andersen, 2007).

We restricted the populations to stars for which standard parallax errors were less that 20% of the quoted parallax. A distance cut of 300pc was also applied. After the distance cuts, the populations contained very few stars with large motion errors. The accuracy of proper motions in HNR is better by a factor of about two than that of Tycho-2 which compared star positions from the Hipparcos satellite with early epoch ground-based astrometry. We used a mean value from HNR and Tycho for proper motion, inversely weighted by the squared quoted error, to obtain the best possible figure. The mean error in transverse velocity is 0.34 kms$^{-1}$, about 1% of the mean transverse velocity, 32.9 kms$^{-1}$. The mean error in radial velocity for the population is 1.3 kms$^{-1}$, for stars also in G-CS the error is 0.87 kms$^{-1}$, and for stars also in Famaey it is 0.26 kms$^{-1}$.

### 2.2 Selection Criteria

Our population of 20 574 stars is obtained by applying the following selection criteria:

(i) Heliocentric distance within 300pc based on HNR parallaxes and parallax error less than 20% of parallax (see section 2.3).

| Population | Stars | proportion | \|pm\| | $\sigma_{|pm|}$ | $\|v_t\|$ | $\sigma_{|vt|}$ |
|---|---|---|---|---|---|---|
| All F&G dwarfs | 9 663 | 100% | 120.4 | 136.6 | 36.1 | 30.9 |
| G-CS F&G dwarfs | 8 148 | 100% | 123.8 | 138.7 | 34.8 | 27.3 |
| All K&M giants | 4 916 | 100% | 61.6 | 61.6 | 35.6 | 25.4 |
| Famaey K&M giants | 1 534 | 100% | 55.1 | 65.1 | 35.3 | 24.7 |
| | | | | | | |
| All F&G dwarfs | 7 296 | 75.5% | 104.6 | 101.8 | 26.2 | 14.2 |
| G-CS F&G dwarfs | 6 190 | 75.9% | 109.3 | 104.5 | 26.6 | 14.3 |
| All K&M giants | 3 489 | 70.1% | 39.3 | 36.4 | 26.0 | 14.3 |
| Famaey K&M giants | 1 098 | 70.6% | 45.7 | 38.4 | 25.8 | 14.7 |

**Table 2:** Top table: Comparison of means and standard deviations of the magnitude of transverse velocity, $v_t$, and magnitude of proper motion, pm, for all F&G dwarfs and G-CS, and for all K&M giants and Famaey. Bottom table: Comparison when the population is restricted to the velocity ellipsoid (section 3). Evidence of selection bias is outweighed by uncertainties due to fast moving stars.

(ii) Radial velocity given in CRVAD-2, GC-S or Famaey and uniquely identified to a Hipparcos catalogue number. CRVAD-2 figures were used by default, as CRVAD-2 gives a weighted mean for stars in Famaey having radial velocities from other sources. We excluded stars for which no radial velocity error was given, or for which the quoted error was greater than $5 \mathrm{km s^{-1}}$.

(iii) The object is either a single star or a spectroscopic binary with a computed mean radial velocity. This criterion is determined from flags provided by G-CS, Famaey, Tycho-2, and CRVAD-2.

(iv) It is usual in statistical analyses of data to eliminate outliers more than three (or fewer) standard deviations from the mean, because outliers tend to have a disproportionately large affect on results. This cannot be done here because the distributions are far from Gaussian and contain a high proportion of fast moving stars. Velocities opposing any error in the mean will be preferentially removed, resulting in a compounded error and leading to non-convergence on iteration of the method. It remains important to remove stars with extreme velocities, especially those with contrary orbits or with orbits excessively inclined to the Galactic plane. A more disperse distribution was found for stars aged over 10 Gyrs. We applied a cut on stars with velocities outside of an ellipsoid,

$$\frac{(U+12)^2}{200^2} + \frac{(V+42)^2}{200^2} + \frac{(W+7)^2}{120^2} < 1,$$

corresponding approximately to a 4 s.d. cut on each axis for the population of old stars, and to over 5 s.d. for the remaining population. This removed 86 stars.

(v) We established subpopulations of 8 098 dwarfs and 6 572 giants and subgiants from five databases of stellar types: *NStars Project* (Gray, et al., 2003 & 2006), *Michigan Catalogue of HD stars*, Vols. 1-5 (Houk & Cowley, 1975; Houk, 1978, 1982, 1988, 1999), *Catalogue of Stellar Spectral Classifications* (Skiff, 2007), *Selected MK Spectral Types* (Jaschek, 1978), and *The Tycho-2 Spectral Type Catalog* (Wright, 2003) by preference in that order.

### 2.3 Parallax Errors

Because parallax distance is measured as an inverse law of parallax angle, errors are not symmetrical and a systematic distance error is introduced (this is a part, but not the main part, of the Lutz-Kelker bias which concerns estimates of absolute magnitude; Lutz & Kelker, 1973, 1974, 1975). For example, for two measurements with 20% error above and below the true parallax, $\pi$, of a given star, the mean parallax distance is

$$R = \left( \frac{1000}{\pi(1-0.2)} + \frac{1000}{\pi(1+0.2)} \right) \div 2 = \frac{1000}{\pi(1-0.2^2)},$$

giving a mean error of +4%. For a Gaussian error distribution with $\sigma = 20\%$ of $\pi$, we calculate an expected systematic error of +1.6% (by numerical solution of the integral). Over 70% of the stars in the population have parallax errors less than 10%. The systematic error goes as the square of the random error and can be estimated at below 1%. We compensated using a pragmatic approximation,

$$R = \frac{1000}{Plx(1 - 0.4(ePlx/Plx)^2)}$$

where *Plx* and *ePlx* are the measured parallax and parallax error given in HNR.

### 2.4 Kinematic Bias

G-CS and Famaey are deemed to be free from kinematic selection bias. The remaining radial velocities in CRVAD-2 are derived from the *General Catalog of Mean Radial Velocities* (Barbier-Brossat and Figon, 2000; hereafter "GCRV") and the *Pulkovo Catalog of Radial Velocities* (Bobylev, et al., 2006). These are compilations from sources some of which may contain a selection bias favoring high proper-motion stars (Binney et al., 1997). Our best methods for determining the LSR exclude high velocity stars.

Concern over a bias towards high proper motions is overstated, since the traditional calculation is fairly insensitive to a kinetic bias with no directional component; a bias toward high velocity stars will increase uncertainty, but the consequent high figures for both $\bar{V}$ and for the asymmetric drift will tend to cancel out in the calculation of $V_0$. It is seen in the analysis that any bias to high motion stars in CRVAD-2 has no impact on results. Binney et al. did not give a statistical analysis for their conclusion, but justified it from a graph (their fig. 2) with a logarithmic scale which exaggerates evidence of bias by two orders of magnitude. Comparison of the statistics for the entire population and for G-CS and Famaey shows little, if any, evidence of selection bias toward high proper motion, (table 2 and table 3), and none when high velocity stars are excluded. Any bias appears to be toward high transverse velocities, not high proper motions. The most likely reason for Binney et al.'s conclusion is statistical errors in their sample resulting from the high proportion of fast moving stars in the population. Skuljan, Hearnshaw and Cottrell (1999) also analysed the claim, finding that "the effect is important only at $v_t > 70\text{-}80 \mathrm{km s^{-1}}$".

A more important consideration is kinetic bias with a directional component. A number of kinematic studies have concentrated on stars in open clusters. These are likely to be over-represented in CRVAD-2. As we will see, the effects of bulk streaming motions dominate over selection bias. To determine the LSR we must find an indicator which is not affected by streams.

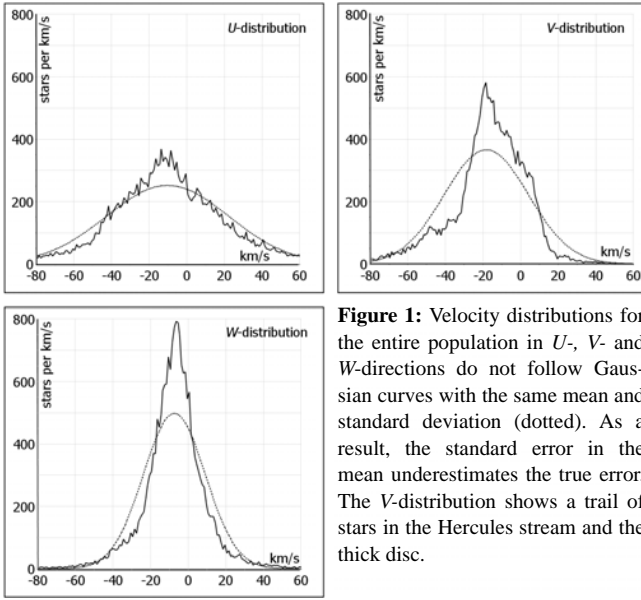

**Figure 1:** Velocity distributions for the entire population in *U*-, *V*- and *W*-directions do not follow Gaussian curves with the same mean and standard deviation (dotted). As a result, the standard error in the mean underestimates the true error. The *V*-distribution shows a trail of stars in the Hercules stream and the thick disc.

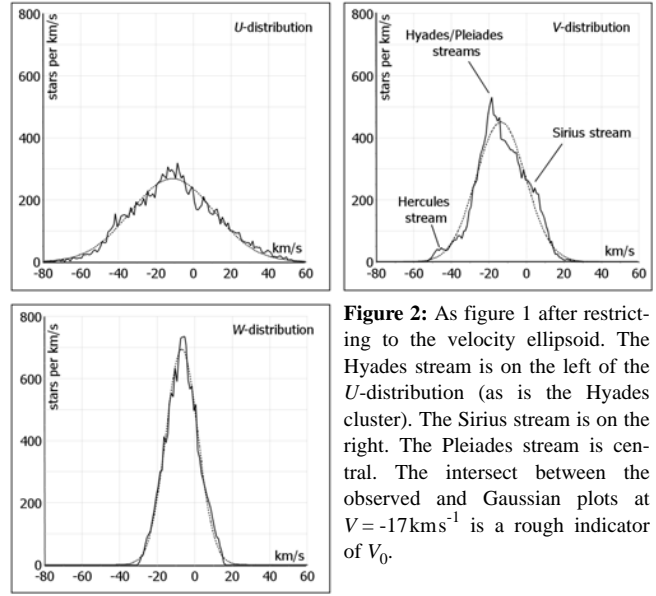

**Figure 2:** As figure 1 after restricting to the velocity ellipsoid. The Hyades stream is on the left of the *U*-distribution (as is the Hyades cluster). The Sirius stream is on the right. The Pleiades stream is central. The intersect between the observed and Gaussian plots at $V = -17 \mathrm{km s}^{-1}$ is a rough indicator of $V_0$.

## 3   Velocity Ellipsoid Cut

For the population of 20 574 stars, the mean velocity is

$$(\overline{U}, \overline{V}, \overline{W}) = (-10.2 \pm 0.2, -18.3 \pm 0.2, -7.4 \pm 0.1) \ \mathrm{km s}^{-1}.$$

Standard deviation is

$$(U_\sigma, V_\sigma, W_\sigma) = (32.6, 22.4, 16.5) \ \mathrm{km s}^{-1}.$$

Calculation of errors is hampered because the velocity distributions are far from Gaussian and contain bulk motions as well as substantial numbers (27.5%) of fast moving stars, including halo stars, thick disc stars, stars produced in high energy events and stars with unusual orbits following collisions or near collisions. As a result, the calculated statistical error understates the true error.

We restricted the populations to mainly thin disc stars with conventional motions in a velocity ellipsoid by fitting the truncated distributions to Gaussian curves. Gaussian fitting is more laborious, but has a number of advantages over truncating to within three standard deviations of the mean. The method converges to a definite central point for the velocity ellipsoid, while cutting at three standard deviations of the mean does not. It is less influenced by moving groups and results in close to Gaussian distributions within an approximately 3σ ellipsoid, particularly for the *U* and *W* distributions. The error in the mean due to the error in the fitted ellipsoid is less than the statistical error in the mean.

Velocity distributions were found using 1 kms⁻¹ bins. There is substantial difference between the observed *U*-, *V*- and *W*-distributions and Gaussian curves with the same mean and standard deviation (figure 1). A high proportion of fast moving stars substantially increases standard deviation so that the Gaussian curve is flatter than the observed distribution. Since fast moving stars are not typical of thin disc motions, and since outliers contribute disproportionately to errors in statistical analysis, it is desirable to remove them from the population. A simple cut, three standard deviations from the mean, cannot be applied because it preferentially removes stars opposing any error in the mean, compounding the error and leading to non-convergence on iteration of the method. Instead we restricted the population to a velocity ellipsoid, and adjusted the ellipsoid to achieve the best least squares fit between the observed distributions of *U*-, *V*- and *W*-velocities within the velocity ellipsoid and Gaussian curves with the same central point and with the same standard deviations (figure 2). It is seen that the distributions for stars inside the ellipsoid are close to Gaussian (bearing in mind the expected asymmetry of the *V*-distribution).

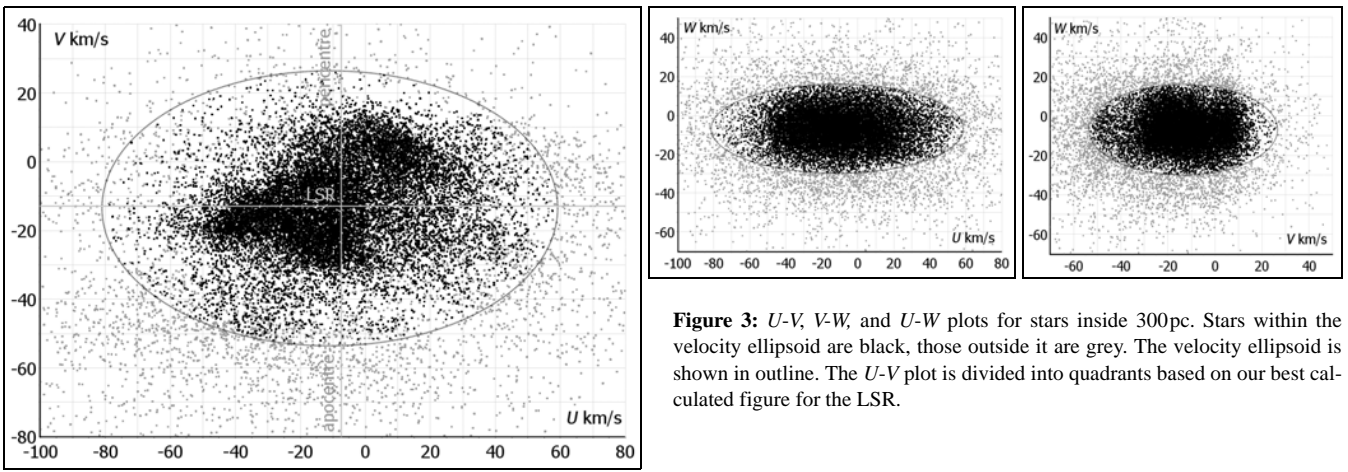

**Figure 3:** *U-V, V-W,* and *U-W* plots for stars inside 300pc. Stars within the velocity ellipsoid are black, those outside it are grey. The velocity ellipsoid is shown in outline. The *U-V* plot is divided into quadrants based on our best calculated figure for the LSR.

Six variables were used for the fit, defining the centre of the ellipsoid, $(U_e, V_e, W_e)$, and the lengths of three semi-axes, $U_r$, $V_r$ and $W_r$. We restricted the population to a velocity ellipsoid,

$$\frac{(U - U_e)^2}{U_r^2} + \frac{(V - V_e)^2}{W_r^2} + \frac{(W + W_e)^2}{W_r^2} < 1 \,,$$

and minimized each of

$$\chi_u^2 = \left( \sum_i (f(u_i) - g(u_i))^2 \right) / N$$

with respect to $u_e$ and $u_r$, where $u = U$, $V$ or $W$, $u_i$ is the central velocity of the $i$th bin, $f(u_i)$ is the number of stars in the $i$th bin, $N$ is the number of stars in the velocity ellipsoid in the current iteration, and

$$g(u) = \frac{N}{u_\sigma \sqrt{2\pi}} \exp\left( \frac{-(u - u_e)^2}{2u_\sigma^2} \right) \,,$$

where $u_\sigma$ is the standard deviation in the current iteration. Initial values for the fitting parameters are not critical (it is natural to start with an ellipsoid centred at the mean for the whole population, and with semi-axes a multiple of standard deviation, figure 1). We iterated each variable by turns, to ensure that a change in one dimension did not alter the optimal fit in another dimension, and terminated the procedure when the centre of the ellipsoid was found to 1 decimal place, and when the semi-axes were found to whole number accuracy. Error bounds were found from a chi² probability distribution after normalizing squared differences.

The centre of the fitted ellipsoid is $(U_e, V_e, W_e) = (-11.2 \pm 0.4$, $-13.6 \pm 0.3$, $-6.9 \pm 0.1$) kms⁻¹. A substantial number, 5 660, of fast moving stars have been discarded, but the ellipsoid contains the bulk of stars with conventional thin disc orbits (figure 3). After restricting the population to

$$\frac{(U + 11.2)^2}{70^2} + \frac{(V + 13.6)^2}{33^2} + \frac{(W + 6.9)^2}{23^2} < 1 \,,$$

the population contained 14 914 stars with mean velocities $(\overline{U}, \overline{V}, \overline{W}) = (-10.0 \pm 0.2, -13.3 \pm 0.1, -6.8 \pm 0.1)$ kms⁻¹ and standard deviations $(U_\sigma, V_\sigma, W_\sigma) = (22.2, 13.2, 8.6)$. The ellipsoid has axes $(U_r, V_r, W_r) = (3.2 U_\sigma, 3.0 V_\sigma, 2.7 W_\sigma)$.

The $V$-distribution shows the expected asymmetry, due to the fact that stars spend longer near apocentre than pericentre, and due to the increased density of stars with lower orbital radius. One may read from this asymmetry that the intersect between the observed and Gaussian plots at $V = -17$kms⁻¹ is a rough indicator of $V_0$.

Although the $U$-distribution is expected to be symmetric while the $V$-distribution is not, there is a greater discrepancy between $\overline{U}$ and $U_e$ than there is between $\overline{V}$ and $V_e$. We will understand this as a consequence of streaming motions. In the absence of a full understanding of the dynamics underlying streams, it is strictly not possible to give an estimate of $V_0$ from either $\overline{V}$ or $U_e$. We have adopted an estimate from $\overline{U}$, in keeping with usual practice, but, without detailed understanding of the causes of asymmetry in the $U$-distribution, one should be cautious of placing much reliance on it.

## 4   Parenago's Discontinuity

We restricted the population to 8 098 dwarfs and binned by colour into 20 bins, each with close to 400 members. We plotted $-\overline{U}$, $-\overline{V}$, $-\overline{W}$ and

$$\sigma_R = \sqrt{\overline{v_R^2}}$$

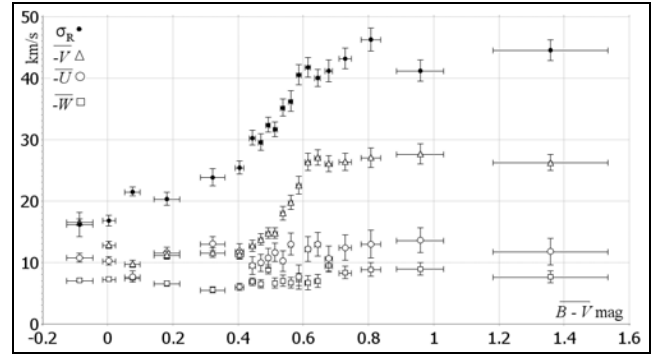

**Figure 4:** Solar motion, $(-\overline{U}, -\overline{V}, -\overline{W})$ and $\sigma_R$, relative to stellar populations binned by colour. The horizontal axis shows $\overline{B - V}$ for each bin with standard error bars.

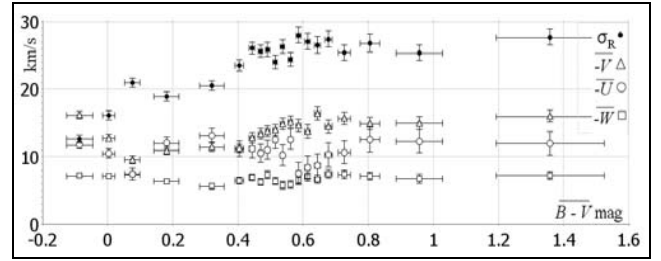

**Figure 5:** As figure 4, for stars within the velocity ellipsoid.

against $\overline{B - V}$ for the bins, where $v_R$ is radial velocity from Sgr A* (figure 4). Parenago's discontinuity (1950) is seen in the plots of $\sigma_R$ (dots) and $-\overline{V}$ (triangles); around $B - V = 0.64$ mag (type G3-4) there is an abrupt change in gradient from a strongly positive value to about zero. Dehnen & Binney suggest that the reason for Parenago's discontinuity is the heating of the disc, scattering processes causing the random velocities of stars to increase with age (e.g., Jenkins 1992). Bluer stars to the left of the discontinuity reflect younger populations, while those to the right of the discontinuity have an age equal to that of the Galactic disc. On the face of it, this would suggest an age of about 10 Gyrs for the Galaxy.

We restricted the population to the velocity ellipsoid (figure 5). The sharp rise above $B - V = 0.4$ mag was eliminated, showing that Parenago's discontinuity is caused by fast moving stars rather than by gradual heating. Quillen & Garnett (2000) found an abrupt, statistically significant, jump in all velocity components at age $9 \pm 1$ Gyrs, corresponding roughly to Parenago's discontinuity, and proposed that the cause might be a galactic merger. It will be only possible to interpret these results by analyzing the age and composition of the various stellar streams (sections 6 & 7). We will find that these effects are associated with the Hercules stream.

## 5   Strömberg's Asymmetric Drift Relation

Strömberg's asymmetric drift equation (e.g., Binney and Tremaine, 1987, 4-34) can be written

$$\overline{V} = V_0 + \frac{\overline{v_R^2}}{D} \,,$$

where $v_R$ is radial velocity from Sgr A*. Using the theoretical value, $D = 110 \pm 7$ kms⁻¹ given by Binney and Tremaine, we obtained, for the population of 20 574 stars,

$$V_0 = 8.2 \pm 0.6 \,,$$

and, for 14 914 stars in the velocity ellipsoid,

$$V_0 = 8.5 \pm 0.3 \,.$$

| Population | Region | Stars | $\overline{U}$ | $\overline{V}$ | $\overline{W}$ | $U_\sigma$ | $V_\sigma$ | $W_\sigma$ | $V_0$ |
|---|---|---|---|---|---|---|---|---|---|
| All stars | All | 19 318 | -10.2±0.2 | -17.8±0.2 | -7.4±0.1 | 32.5 | 21.2 | 15.8 | 8.2±0.6 |
| All F&G dwarfs | All | 8 124 | -10.7±0.4 | -18.1±0.2 | -7.4±0.2 | 33.7 | 21.4 | 16.6 | 7.8±0.7 |
| G-CS F&G dwarfs | All | 6 691 | -10.4±0.4 | -18.1±0.3 | -7.4±0.2 | 33.2 | 21.2 | 16.6 | 8.1±0.7 |
| All K&M giants | All | 4 815 | -8.9±0.5 | -20.0±0.4 | -7.5±0.3 | 34.3 | 22.3 | 17.4 | 9.3±0.8 |
| Famaey K&M giants | All | 3 295 | -7.9±0.6 | -20.4±0.4 | -7.7±0.3 | 34.2 | 22.3 | 17.4 | 9.8±0.8 |
| | | | | | | | | | |
| All stars | Ellip. | 14 914 | -9.9±0.2 | -13.2±0.1 | -6.8±0.2 | 22.8 | 13.4 | 8.7 | 8.5±0.3 |
| All F&G dwarfs | Ellip. | 5 630 | -10.4±0.3 | -13.8±0.2 | -6.7±0.1 | 23.9 | 13.6 | 9.2 | 8.6±0.4 |
| G-CS F&G dwarfs | Ellip. | 4 610 | -10.4±0.4 | -14.1±0.2 | -6.7±0.1 | 24.3 | 13.9 | 9.4 | 8.7±0.4 |
| All K&M giants | Ellip. | 3 071 | -8.4±0.4 | -14.3±0.3 | -6.7±0.2 | 24.4 | 14.2 | 9.3 | 8.9±0.4 |
| Famaey K&M giants | Ellip. | 2 077 | -7.7±0.5 | -14.7±0.3 | -6.6±0.2 | 24.1 | 13.9 | 9.3 | 9.4±0.5 |

**Table 3:** Calculated mean velocities and standard deviations in kms$^{-1}$ for different populations, together with the value of $V_0$ corrected for the asymmetric drift using the theoretical value of $D$. The restriction to G-CS and to Famaey shows little difference from the distributions for F&G dwarfs or K&M giants. Statistical errors are likely to be understated for the full population because of the non-Gaussian nature of the distribution. The true error in $V_0$ is dominated by streaming bias.

The major part of the error is due to uncertainty in $D$, but the error for the entire population may be understated because the distribution is not Gaussian. The reduced error for stars in the velocity ellipsoid is due both to the reduction in uncertainty in $\overline{V}$ and to a reduced value of $\overline{v_R^2}$. We repeated the calculation for G-CS and Famaey to compare the results of these kinematically unbiased populations with the full populations of F&G dwarfs and K&M giants (table 3). Comparison between the figures indicates that statistical errors outweigh possible selection bias.

Strömberg's asymmetric drift equation gives a linear relation between $V_0$ and $\overline{v_R^2}$. It is generally thought that, in principle, one can plot a line of best fit, and read $V_0$ from the intersect with the vertical axis (e.g. Dehnen & Binney, 1998). Figure 6 shows the regression for dwarfs binned by colour. The two bluest bins represent populations of young stars which may be expected a kinematic behaviour different from the background. When they are excluded the intercept is $V_0 = 4.1 \pm 1.5$ kms$^{-1}$. Restriction to the velocity ellipsoid reduces both $-\overline{V}$ and $\overline{v_R^2}$, and leads to a consistent result but the quality of the regression is poor. There is no correlation for the fourteen bins with $B - V > 0.427$ mag (i.e. later than ~F3-4), indicating that the correlation has been produced by a population of fast moving stars, not by progressive changes in a well-mixed distribution.

The method strictly requires a kinematically unbiased population. We repeated the exercise for G-CS and Famaey. We found no useful correlation for the giants. For G-CS the intercept was found at $V_0 = 3.8 \pm 1.5$ kms$^{-1}$. The two reddest bins in G-CS lie outside of the line of the rest of the population. When they are excluded the intercept drops to $V_0 = 1.1 \pm 1$ kms$^{-1}$. The results of these calculations show larger than expected errors. Examination of figures 1 and 3 suggests that they are unreasonable. $V_0 < 6$ kms$^{-1}$ would imply that over 70% of the population trail the LSR. By a rough estimate, for a well-mixed distribution, the radial distance of a typical orbit to pericentre would be about half that to apocentre. Since orbital velocity at apocentre would then be half that at pericentre (by conservation of angular momentum and the flatness of the rotation curve), and since orbital velocities are distributed between a minimum at apocentre and maximum at pericentre, velocity dispersion would be greater than the observed dispersion by an approximate factor of four.

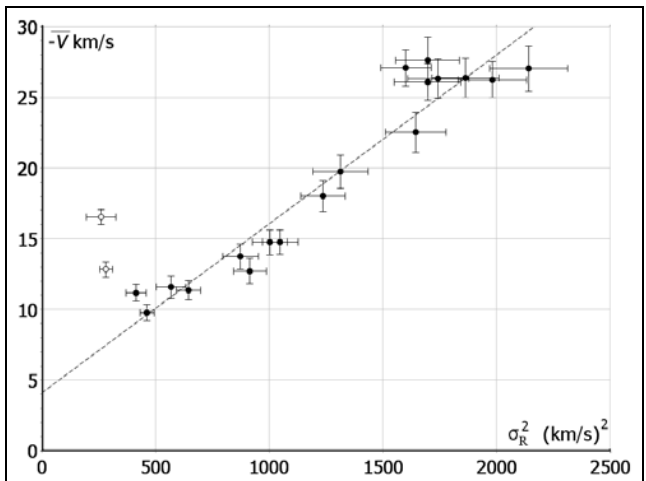

**Figure 6:** $-\overline{V}$ against $\sigma_R^2 = \overline{v_R^2}$ binned by colour. The line of regression is shown, excluding $B - V < 0.057$ mag (open circles).

## 6 Bulk Streaming Motions

The existence of stellar streams was first established from astronomical investigations dating as far back as 1869 (Eggen, 1958). They were thought to consist of previously clustered coeval stars that have been gradually dispersed by the dynamic processes of tidal forces, differential galactic rotation, and encounters with other stars. Increasingly comprehensive star catalogues published in the 1950's opened the way for more thorough analyses. Beginning in 1958, O.J. Eggen produced a series of seminal studies of stellar streams using RA:DE proper motion ratios in conjunction with radial velocities. The results of Eggen's investigations realized significantly increased membership counts and spatial extents of stellar streams. Eggen hypothesized a more protracted process of dissolution for star clusters. In Eggen's scenario, as star clusters dissolve during their journeys around the Galaxy, they are stretched into tube-like formations which were subsequently called superclusters.

The investigation of stellar streams received a major boost with the arrival of the precision astrometry afforded by the Hipparcos mission. Dehnen (1998), using transverse velocities derived from Hipparcos, produced maps of the local stellar velocity distribution showing that streams contain a significant proportion of late type stars. A wide range of stellar ages was identified within superclusters, challenging Eggen's hypothesis of common origin (e.g., Chereul et al., 1998, 1999). Building on the theoretical groundwork of Kalnajs (1991), Dehnen (1999) described a mechanism in which the outer Lindblad resonance of the Galactic bar could elongate closed orbits (in two modes) near the solar circle. Dehnen (1999, 2000) called its predicted effect "resonant scattering" and identified a signature in the Hipparcos data. This mechanism appears to work for the Hercules stream (Fux, 2001).

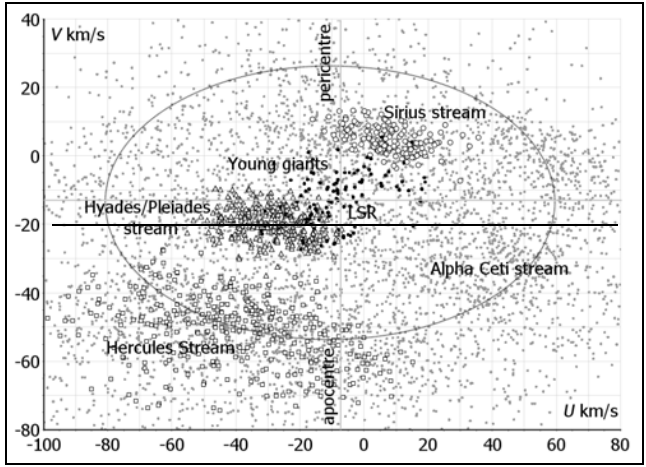

**Figure 7:** *U-V* plot showing groups identified by Famaey et al. (2005) (Sirius stream: circles, young giants: dots, Hyades/Pleiades: triangles, Hercules stream: squares). These represent only a proportion of the true membership of the streams. The calculated position of the LSR is shown for clarity only.

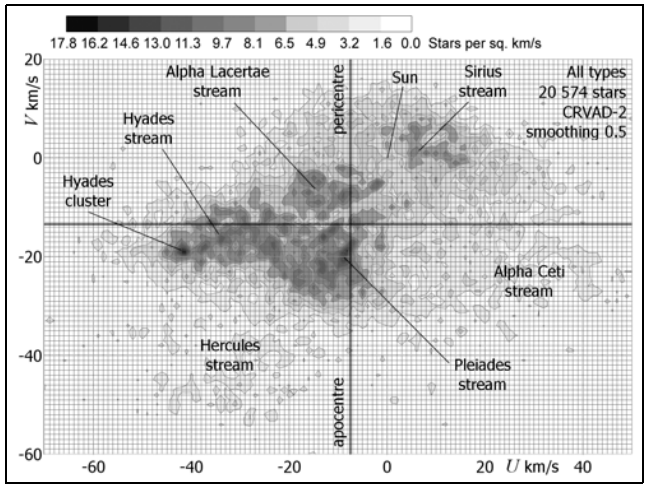

**Figure 8:** The distribution of *U*- and *V*-velocities using Gaussian smoothing with a standard deviation of 0.5kms$^{-1}$, showing the Hyades, Pleiades, Sirius, Alpha Ceti, and Hercules streams.

**Figure 9:** The variation of the velocity distribution with respect to stellar type. Density contours are relative to the peak density for each plot. The optimal choice of smoothing parameter is subjective and depends on stellar density. Too high a value will suppress details of structure, while too low a value may cause random fluctuations to appear as structure.

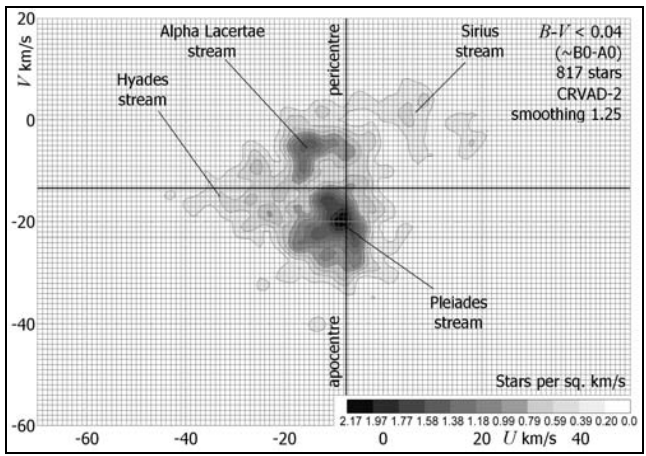

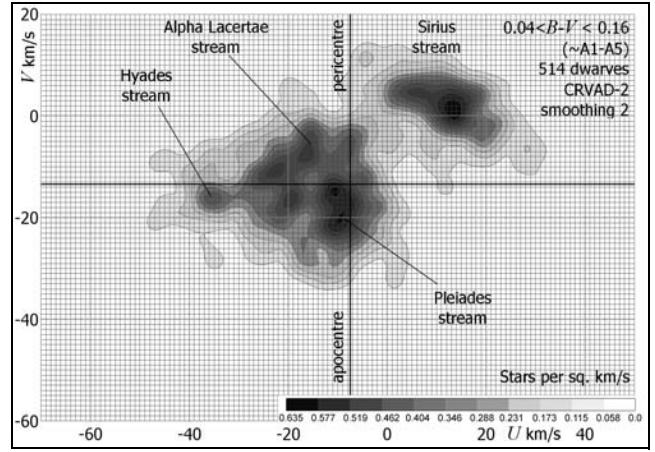

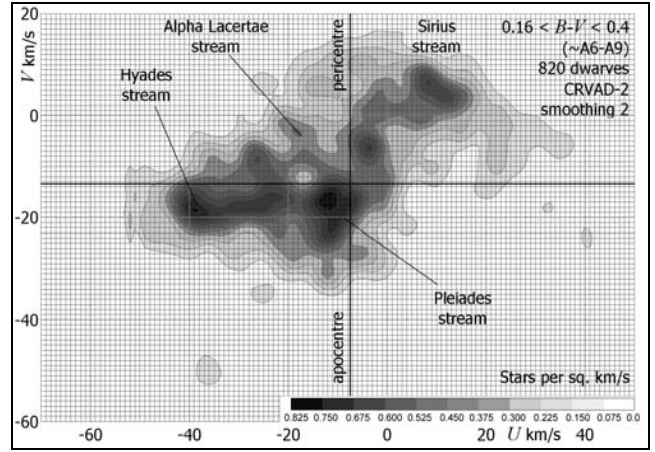

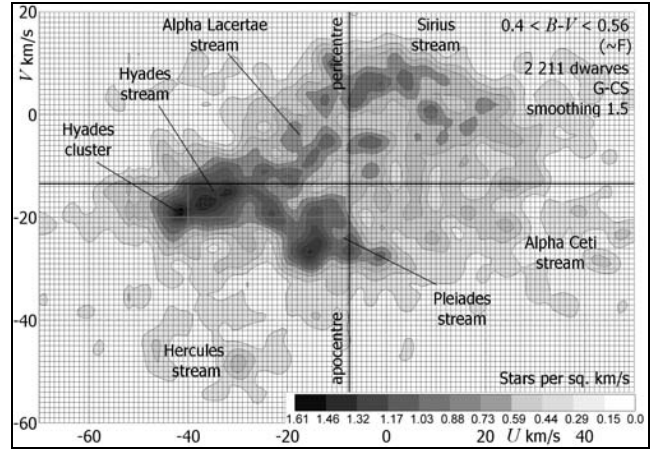

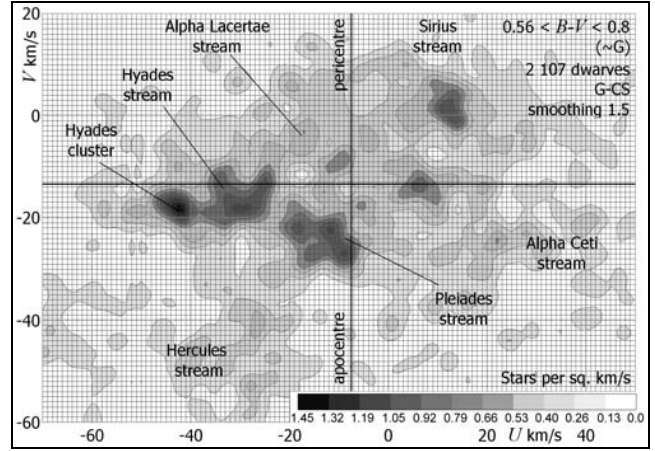

Streams are not necessarily all formed in the same way and the search for other types of dynamical mechanisms to account for the Sirius, Hyades and Pleiades streams is ongoing. Candidates include migrations of resonant islands (Sridhar & Touma, 1996; Dehnen, 1998) and transient spiral waves (De Simone et al., 2004; Famaey et al., 2005) in which streams originate from perturbations in the gravitational potential associated with spiral structure.

Famaey et al. (2005) described six kinematic groups: three streams, Hyades/Pleiades, Sirius and Hercules, a group of young giants, high velocity stars and a smooth background distribution (figure 7). We smoothed the velocity distribution by replacing each discrete point with a two dimensional Gaussian function and finding the sum. A standard deviation of 0.5 gave a clear contour plot (figure 8). We distinguish the Hyades and Pleiades streams, since the velocity distributions shows separate peaks, and, as we will see, these streams contain different distributions of stellar types and ages. There is a large and well dispersed stream centred at $(U, V) \approx (25, -23)\,\mathrm{km\,s^{-1}}$. This estimate is in good agreement with Chakrabarty (2007) who identified a clump in the velocity distributions at $(U, V) \approx (20, -20)\,\mathrm{km\,s^{-1}}$. We have called it the Alpha Ceti stream, after the brightest star we identified with stream motions. We also distinguish the Alpha Lacertae stream, which contains young stars, but has distinct motion from the Pleiades stream. Famaey's young giants are mainly in this stream.

Figure 9 shows changes in the velocity distribution with respect to stellar type. More detailed information about the structure of streams was also gleaned using narrower colour bands. There are few candidates for the Sirius and Hyades streams earlier than B7, and these appear to be part of the distribution for stars with young kinematics. The earliest indication of the Sirius stream as a distinct distribution is for stars of type B8, corresponding to an age of about 300 Myrs, and for the Hyades stream for type B9, an age of about 400 Myrs. The Hercules and Alpha Ceti streams are both apparent at type F0, corresponding to an age of about 2½ Gyrs, with too few candidates of earlier types to draw conclusions.

Comparison of the distributions in figure 9 with figure 4 shows that the values of $-\overline{V}$ and $\sigma_R$ for different colour depends heavily on the structure of the velocity distribution. The bluest stars reflect recent star formation in the Pleiades and Alpha Lacertae streams, leading to a low value of $\sigma_R$. For $0.04 \leq B - V < 0.16$, the velocity distribution is concentrated in the Pleiades and Sirius streams, resulting in a rise in $\sigma_R$ and the minimum of $-\overline{V}$ seen in figure 4. For $0.16 \leq B - V < 0.4$ the Hyades stream begins to dominate, resulting in the increased values of both $-\overline{V}$ and $\sigma_R$. Finally, for F & G dwarfs, the Hercules and Alpha Ceti streams cause further increases in $-\overline{V}$ and $\sigma_R$, seen leading up to Parenago's discontinuity in figure 4, and the reduced importance of the Sirius stream also causes $-\overline{V}$ to increase.

In conclusion, dependencies on colour show that total stream membership is underestimated in Famaey's figures. The increasing value of $-\overline{V}$ with respect to colour is seen in figures 9b-e, and depends heavily on stream composition. The slope of the regression in figure 6 is dictated by the structure of the velocity distribution, and has no bearing on Strömberg's asymmetric drift equation. In fact the argument by binning is based on a fallacy. There is no evidence that if the population were well-mixed there would be a correlation between $-\overline{V}$ and $v_R^2$. The correlation in figure 6 dem-onstrates structure invalidating the calculation of the asymmetric drift. Streams violate the assumption of a well-mixed distribution; we cannot determine the value of $D$ empirically through binning the population by type or by colour.

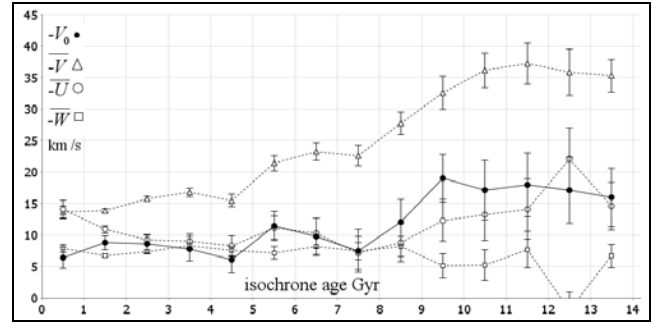

**Figure 10:** Values of $-\overline{U}$, $-\overline{V}$, $-\overline{W}$ and $V_0$ for populations in 1 Gyr bins, using $D = 110 \pm 20$. The midpoint of the bin is shown. Connecting lines are drawn for clarity. The last bin contains all stars given as $> 13$ Gyrs, and is likely to contain a number of very young stars.

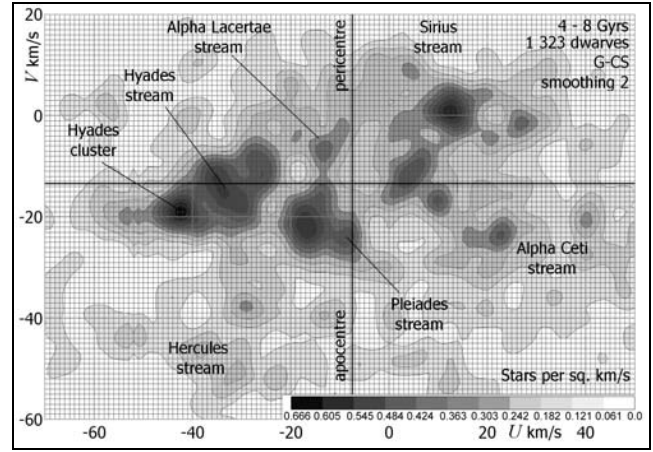

**Figure 11:** The $U$-$V$ density for stars aged between 4 and 8 Gyrs. Nine members of the Hyades cluster are contained in this group.

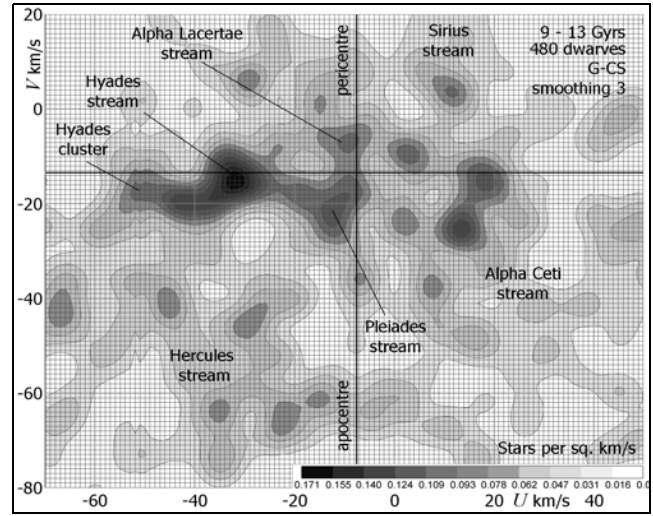

**Figure 12:** The $U$-$V$ density for stars aged between 9 and 13 Gyrs. Four members of the Hyades cluster are contained in this group.

Gas clouds from which stars form are expected to have a more nearly circular motion than the norm. Recently formed stars can be expected to have a more nearly circular motion than the population as a whole. The Pleiades stream is seen in figures 9a-c to have a strong peak in the vicinity of $(U, V) = (-10 \pm 2, -16 \pm 3)\,\mathrm{km\,s^{-1}}$. Although this does not give a precise estimate of the LSR, it may be regarded as a rough guide to the region of velocity space in which the LSR is to be found.

## 7  Old Stars

It might be expected that a population of old stars will be sufficiently well-mixed to carry out a calculation of the asymmetric drift. G-CS II isochrone ages appear to be at least broadly reasonable, and were supported by an H-R diagram showing age bands in accordance with theory, to be reported in a paper in preparation. An explanation for the number of old stars in the Hyades cluster has been found and will also be reported. There are known problems with isochrone aging for very young stars; we found that a substantial number of stars with young kinematics had been assigned ages greater than 13 Gyrs. We binned G-CS into age groups of 1 Gyr, and calculated $-\overline{U}, -\overline{V}, -\overline{W}$ and $V_0$ for each population using $D = 110 \pm 20$, which gives a measure of agreement for the younger populations with our previous values (figure 10). $-\overline{V}$ rises dramatically with age and there is a sudden shift in the calculated value of $V_0$ at 9 Gyrs. However, plots of the velocity distribution show that a well-mixed population has not been found (figures 11 & 12). The appearance of groups with young kinematics in these plots may be accounted for by incorrectly aged stars, but the overall pattern appears significant. The changes in $-\overline{V}$ and $V_0$ are caused by the reduced importance of the Sirius stream and the increased prominence of the Hercules and Alpha Ceti streams in the bins of age greater than 9 Gyrs. It is possible to identify that these are streams of older stars and that the rise in $-\overline{V}$ and $\sigma_R$ with $B - V$ in figure 4 is due to the influence of these streams, not gradual heating.

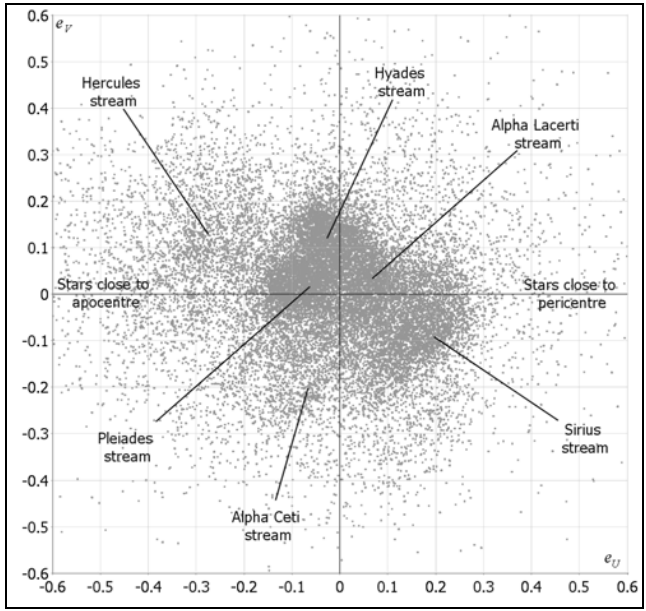

**Figure 13:** The distribution of eccentricity vectors is not homogeneous in the *U-V* plane. The plot is based on our best estimate of the LSR.

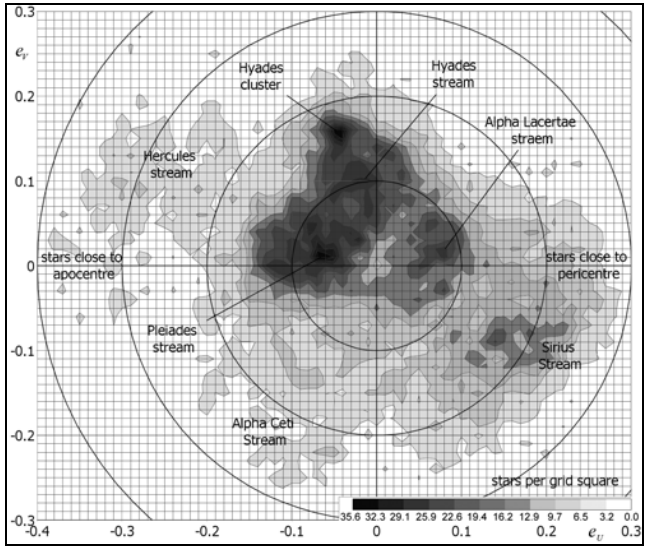

**Figure 14:** Contour of the density of the eccentricity distribution. The Hercules stream has eccentricities up to ~0.3 and orbits approaching apocentre. The Sirius and Alpha Ceti streams have eccentricities ~0.1-0.25 approaching pericentre. The Hyades stream has eccentricities below ~0.2 approaching apocentre. The Pleiades stream has eccentricities below ~ 0.1 close to apocentre.

## 8  The Eccentricity Distribution

For an elliptical orbit the eccentricity vector is defined as the vector pointing toward pericentre and with magnitude equal to the orbit's scalar eccentricity, $e$. It is given by

$$e = \frac{|v|^2 r}{\mu} - \frac{(r \cdot v) v}{\mu} - \frac{r}{|r|},$$

where $v$ is the velocity vector, $r$ is the position vector, and $\mu = GM$ is the standard gravitational parameter for orbits about a mass, $M$. For a Keplerian orbit the eccentricity vector is a constant of the motion. Stellar orbits are not elliptical because mass is distributed in the disc and in the halo. In addition, the orbit will oscillate in the $W$-direction due to the gravitational attraction of the disc, rather than being truly planar. The eccentricity vector is expected to precess from both these causes, such that the orbit is a rosette. Nonetheless, the orbit will approximate an ellipse at each part of its motion, and the eccentricity vector remains a useful measure. It is equivalent up to a scale factor with the Laplace-Runge-Lenz vector which is also used in the study of perturbations to elliptical orbits. Over time the eccentricity vectors of different stars are expected to precess at different rates. It is usually assumed that, in time, an equilibrium state will be attained in which the distribution is well-mixed.

In a well-mixed population the eccentricity vectors will be spread smoothly in all directions, with an overdensity at apocentre and underdensity at pericentre, because of the increased orbital velocity at pericentre and because stars at apocentre come from a denser population nearer the Galactic centre. This is not seen in a plot of the distribution of eccentricity vectors (figure 13). In practice stellar streams are found in which the eccentricity vectors are concentrated at particular values. We smoothed the eccentricity distribution by replacing each discrete point with a two dimensional Gaussian function and finding the sum. Standard deviation, $\sigma$, is used as a smoothing parameter. A standard deviation of 0.005 gave a clear contour plot (figure 14) showing that mixing is poor. The structure of the distribution is largely determined from streaming motions.

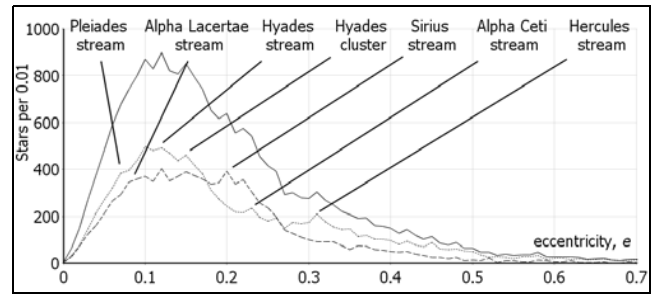

**Figure 15:** Eccentricity distribution (based on the LSR found in this paper) for the entire population, for stars closer to apocentre (dots) and stars closer to pericentre (dashes), as defined by position with respect to the semi-latus rectum. The number of stars closer to apocentre is expected to outweigh the number closer to pericentre, by at most about 20% for $e = 0.1$, and more for larger eccentricities.

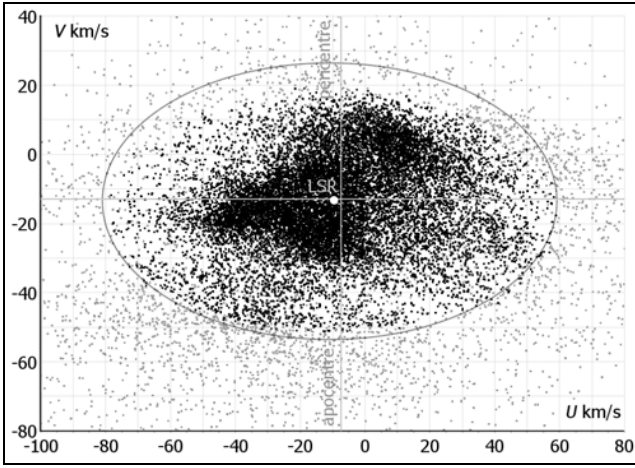

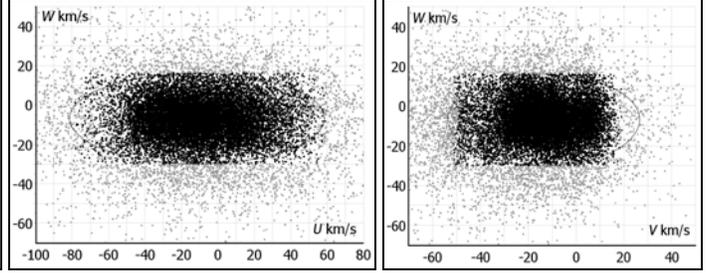

**Figure 17:** *U-V*, *V-W*, and *U-W* velocity plots for the velocity cylinder, stars with eccentricities less than 0.32 and $|W + 6.8| < 22$ km s$^{-1}$ (black), compared to the velocity ellipsoid, shown in outline, and the remaining population (grey). The *U-V* plot is divided into quadrants based on our best calculated figure for the LSR. The estimate of the LSR by the eccentricity cut is shown by a white dot, and is offset from the centre of the cylinder by about 5 km s$^{-1}$ in the *V*-direction.

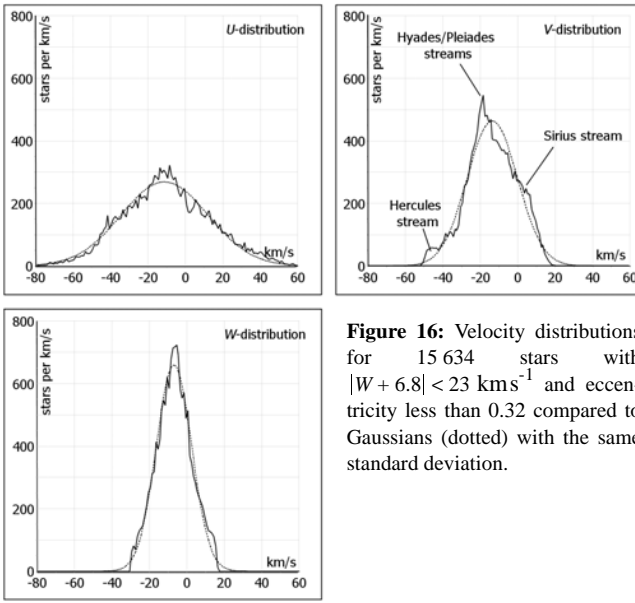

**Figure 16:** Velocity distributions for 15 634 stars with $|W + 6.8| < 23$ km s$^{-1}$ and eccentricity less than 0.32 compared to Gaussians (dotted) with the same standard deviation.

## 9   Analysis with an Eccentricity Cut

Ignoring the possibility of perturbations to the galactic plane, motions of thin disk stars in the *W*-direction may be treated as a low amplitude oscillation due to the gravity of the disc, and as independent of orbital motion in the *U-V* plane. A better representation of the velocity distribution of the thin disc may be found by discarding the velocity ellipsoid, and instead restricting by eccentricity in the *U-V* plane and by restricting *W*-velocity to within a range centred on $W_0$. For any given value for the LSR, and any $e_0$ with $0 < e_0 < 1$, one may find a population of stars with eccentricity in the Galactic plane less than $e_0$. We found values for $(U_0, V_0, W_0)$ and $e_0$ by fit-ting the truncated distributions to Gaussians by adapting the method used to find the velocity ellipsoid in section 3. The advantages of this method are: a) It finds an estimate of the LSR directly, without a separate correction for asymmetric drift. b) Cutting on eccentricity better represents the kinematic properties of stars in the thin disc c) It uses fewer fitting parameters, so is less prone to statistical fluctuations. d) Although streams are clearly apparent in the eccentricity distributions for stars closer to apocentre and stars closer to pericentre, as defined by position with respect to the semi-latus rectum, the full distribution has a smoother form (figure 15). The assumption that streaming motions will largely cancel therefore is less doubtful.

Four (independent) variables were used in the fit. We minimized the combined sums of *U* and *V* squared differences to find $V_0$ and the eccentricity bound, $e_0$ (one may also minimize either *U* or *V* squared differences, or some other linear combination, leading to a small statistical variation in the result). $U_0$ was set to $-\overline{U}$ for the previous iteration, and was found to converge. We minimized *W* squared differences to find the bound on *W*-velocity and $W_0$, which converged to a value close to the mean. The fit for the resulting population is shown in figure 16. The velocity ellipsoid is now replaced with an oval cylinder containing 15 634 stars with eccentricities less than 0.32 and with $|W + 6.8| < 23$ km s$^{-1}$ and giving an estimate of the LSR:

$(U_0, V_0, W_0) = (9.8 \pm 0.2 , 13.2 \pm 1.3 , 6.8 \pm 0.1 )$ km s$^{-1}$.

The statistical error in $\overline{V}$ is less than that in $V_0$:

$\overline{V} = -13.9 \pm 0.4$ km s$^{-1}$.

The standard deviation is

$(U_\sigma, V_\sigma, W_\sigma) = (23.2, 13.5, 9.5)$ km s$^{-1}$.

Despite containing substantially more stars, the cylinder is more compact than the ellipsoid. For the observed local velocity distribution, the centre of the velocity ellipsoid is unexpectedly close to the LSR, but the asymmetric drift is seen in figure 17. The centre of the oval in the *U-V* plane is offset by about 5 km s$^{-1}$ in the *V*-direction from the estimate of the LSR. This shows an effect of streaming bias; if the population were well-mixed, the centre of the velocity ellipsoid would be offset by a similar amount.

## 10   Circular Orbits

Disc heating is the process by which scattering events cause the random velocities of stars to increase with age (e.g., Jenkins, 1992). Even in thermal equilibrium, one would expect a modal value of random peculiar velocity denoting disc temperature. Circular motion represents an absolute zero temperature and can therefore be expected to be rare for mature orbits. As a result, the distribution in velocity space can be expected to have a minimum at circular motion. In practice the situation is not so simple. Figure 8 shows a deep trough in the vicinity of $V = -12 \text{km s}^{-1}$, containing a number of minima. These do not give a precise estimate of the LSR.

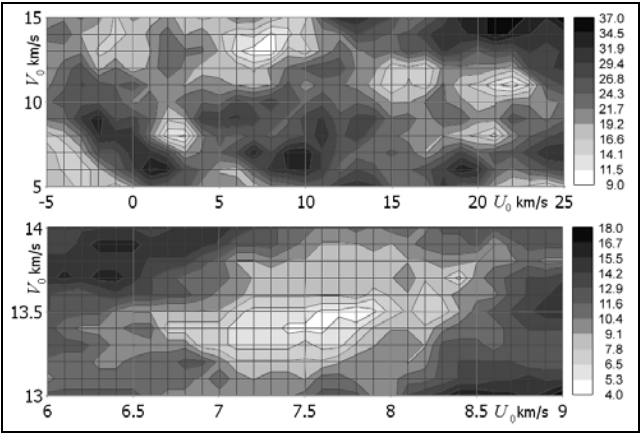

**Figure 18:** The number of stars with eccentricity less than 0.01 for different values of $(U_0, V_0)$, in multiples of $1 \text{km s}^{-1}$, and focusing on the least minimum in multiples of $0.1 \text{km s}^{-1}$. Stars with $B - V < 0.3$ mag are excluded. The position of the minimum gives our best estimate for the LSR, $(U_0, V_0) = (7.5 \pm 1.0 , 13.5 \pm 0.3) \text{km s}^{-1}$.

We plotted the number of stars with eccentricity less than 0.01 for a range of values of $U_0$ and $V_0$ (figure 18). Eliminating the youngest population of blue stars causes the minima to both get both deeper and wider, in keeping with the notion that they are caused by a heating effect. $B - V < 0.3$ mag gave deep minima, but for $B - V < 0.4$ mag the minima become broader, and their positions less precise. The strongest candidate for the minimum at the LSR found at $(U, V) = (7.5 \pm 1.0 , 13.5 \pm 0.3) \text{km s}^{-1}$. Other candidates were finally rejected after analysis of the circular speed curve (section 11). This estimate is independent of kinetic bias due to streams or selection, and gives our best estimate of the LSR.

## 11  The Circular Speed Curve

We restricted the population to stars close to orbital extrema, having $|U - U_0| < 7 \text{km s}^{-1}$, for a range of values of $U_0$. We plotted the transverse orbital velocity against distance to SgrA*, based on an adopted transverse solar velocity of $225 \text{km s}^{-1}$. On the assumption that Sgr A* is stationary at the Galactic barycentre, the proper motion of Sgr A* determined by Reid and Brunthaller (2004) implies a distance to the Galactic centre of $R_0 = 7.4 \pm 0.04 \text{kpc}$, consistent with recent determinations (Reid, 1993; Nishiyama et al., 2006; Bica et al., 2006; Eisenhauer et al., 2005; Layden et al., 1996). For values of $U_0 = 7.5 \pm 2.5 \text{km s}^{-1}$, the scatter plot of the distribution (figure 19) divides clearly into two parts, with a less densely populated band of stars which we believe to be on near circular orbits.

Young stars have velocities dependent on the kinematics of the gas clouds from which they are formed, and are suspected to have motions close to the LSR. We removed stars with $B - V < 0.3$ mag. This increased the visual clarity of the split. There is a noticeable degradation in the quality of the split outside the range $U_0 = 7.5 \pm 0.5 \text{km s}^{-1}$. Outside of $U_0 = 7.5 \pm 2.5 \text{km s}^{-1}$ the split was barely visible (at this dot size). We believe that this confirms the

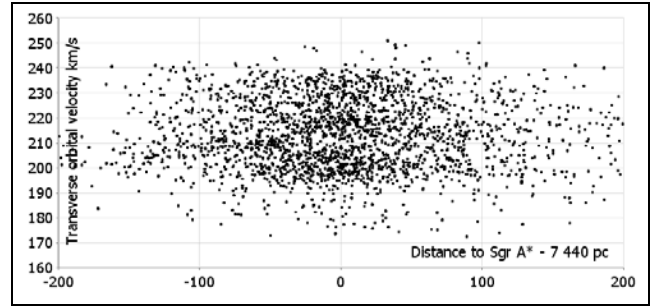

**Figure 19:** Transverse orbital velocities of 2 350 stars with $|U + 7.5| < 7 \text{km s}^{-1}$ plotted against distance to Sgr A*. The circular speed curve is seen in the dearth of stars on circular orbits.

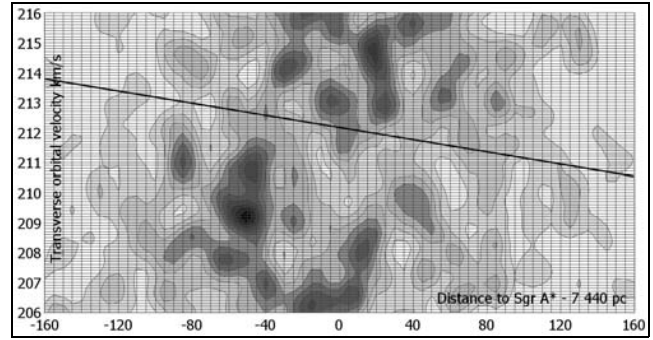

**Figure 20:** The transverse velocity distribution of stars close to circular motion within $206 < V_{\text{T}} < 216 \text{km s}^{-1}$, excluding stars with $B - V < 0.3$ mag, using Gaussian smoothing with parameters $\sigma_v = 0.4$ $\sigma_R = 5$. The line of regression through the minima at constant distance is also shown. The correlation is significant at 99%.

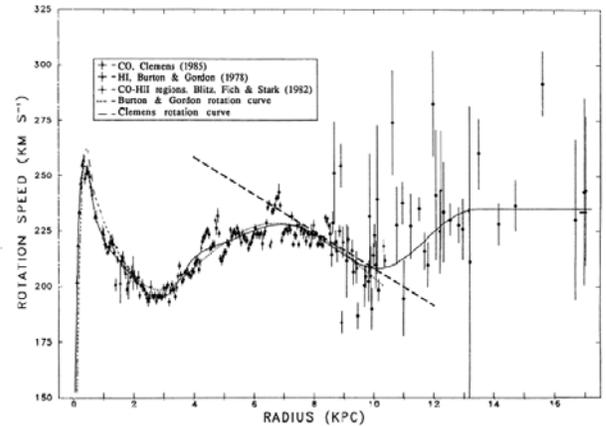

**Figure 21:** The Milky Way rotation curve from CO and HI (figure from Combes, 1991, with permission), with superposed the gradient (dashed) from local stars adjusted to $R_0 = 8.5 \text{kpc}$, $V_0 = 220 \text{km s}^{-1}$ as used by Combes.

figure $U_0 = 7.5 \pm 1.0 \text{km s}^{-1}$ found by calculating circular orbits (section 10) and eliminates the alternate minima.

We restricted distances to 160pc (because of the low population outside this range) and used Gaussian smoothing to find the density of the frequency distribution (figure 20). The trough was well displayed for a range of smoothing parameters, $0.3 \le \sigma_v \le 0.6$, $3 \le \sigma_R \le 6$. Too large values of the smoothing parameters cause interference between peaks in the difference and the minima, while too small values broke up the distribution excessively. We used regression to find a line of best fit to the minima at given distance from SgrA*, and found the intercept at $211.5 \pm 0.5\,\mathrm{km\,s^{-1}}$, corresponding to $V_0 = 13.5 \pm 0.5\,\mathrm{km\,s^{-1}}$, and a slope of $-9.3 \pm 0.9\,\mathrm{km\,s^{-1}\,kpc^{-1}}$, giving a close match with the local slope of the Milky Way rotation curve from CO and HI given by Combes (1991; figure 21).

There is some uncertainty in the slope on account of the short distance for which the population is sufficiently dense to find a meaningful minimum in the trough. Moving groups with close to circular motion also increase uncertainty. However, the existence of the trough in the distribution is significant. The method for calculating both the LSR and the circular speed curve will become more valuable when data from Gaia becomes available. It will be potentially be possible to extend the analysis to a much larger region of space, perhaps even to trace the circular speed curve to near the centre of the Galaxy, and a similar distance outward from the Sun where current methods are problematic.

## 12 Conclusions

The velocity distribution of local stars is highly structured, and heavily biased towards membership of six major streams. The distribution of stream membership contains dependencies on both colour and age. These dependencies, not Strömberg's asymmetric drift relation, are responsible for the correlations between $\overline{V}$ and $\sigma_R^2$ from which the asymmetric drift is usually calculated, and invalidate the usual calculation of the LSR, for which a well-mixed distribution is required. The origin of Parenago's discontinuity is the existence of fast moving streams of older stars, not continuous heating of the disc.

Using statistical analysis, it is only possible to put a least bound on stream membership, not to identify a particular background population to which standard determinations of the LSR might be applied. Streams and moving groups appear more prominently using the HNR than with the less accurate Hipparcos 1997 catalogue. Factors such as the dependency of the distribution on colour and age, and the lack of stars in certain regions of velocity space, have lead us to believe that there is no background distribution as such, but there is no simple way to quantify this conclusion.

We found alternative indicators by examining the properties of the velocity distributions. We believe the best indicators are based on an observed (and unanticipated) minimum in the distribution which we believe represents circular motion. We have accounted for this minimum as a consequence of heating of the disc. A more rigorous argument requires detailed analysis of the relationship between streams and spiral structure, which is the subject of Francis & Anderson (2009). The analysis supports the notion that the minimum represents circular motion. An important strength of the method is that it is unaffected by the dynamical properties of moving groups and streams with non-circular motions.

We found a good measure of agreement between methods (table 4). Our best estimate of the LSR is $(U_0, V_0, W_0) = (7.5 \pm 1.0, 13.5 \pm 0.3, 6.8 \pm 0.1)\,\mathrm{km\,s^{-1}}$. $W_0$ is found from the mean after restricting the population by Gaussian fitting. $U_0$ and $V_0$ are found from the low frequency of stars in orbits with eccentricity less than 0.01, supported by the observed trough in the distribution for stars close to orbital extrema, from which we have derived the slope of the circular speed curve on the assumption that the trough corresponds to circular motion. This estimate is independent of kinetic bias due to streams or selection. These figures are consistent with the supposition that SgrA* is stationary at the Galactic barycentre at a distance of $7.4 \pm 0.2\,\mathrm{kpc}$ and a Solar transverse orbital velocity of $225 \pm 9\,\mathrm{km\,s^{-1}}$.

We calculated the local slope of the circular speed curve found from the low frequency of circular orbits. An unbiased estimate was found by restricting the population to stars close to orbital extrema. For 2 350 stars with $|U + 7.5| < 7\,\mathrm{km\,s^{-1}}$ and $B - V > 0.3$ mag the correlation is significant at 99%, and the slope is $-9.3 \pm 0.9\,\mathrm{km\,s^{-1}\,kpc^{-1}}$, in agreement with the Galactic rotation curve found from CO and HI. A slope of this magnitude suggests that the local mass distribution does not reflect a smooth global distribution of dark matter. Data from Gaia will make it possible to use this low density to trace the circular speed curve over much greater distances.

**Data**
The compiled data used in this paper can be downloaded from
http://data.rqgravity.net/lsr/